\documentstyle[12pt,epsfig]{article}
\date{}
\begin{document}
\title{{\LARGE\sf Nature of Ground State Incongruence in Two-Dimensional Spin Glasses
}}  
\author{
{\bf C. M. Newman}\thanks{Partially supported by the 
National Science Foundation under grant DMS-98-02310.}\\
{\small \tt newman\,@\,cims.nyu.edu}\\
{\small \sl Courant Institute of Mathematical Sciences}\\
{\small \sl New York University}\\
{\small \sl New York, NY 10012, USA}
\and
{\bf D. L. Stein}\thanks{Partially supported by the 
National Science Foundation under grant DMS-98-02153.}\\
{\small \tt dls\,@\,physics.arizona.edu}\\
{\small \sl Depts.\ of Physics and Mathematics}\\
{\small \sl University of Arizona}\\
{\small \sl Tucson, AZ 85721, USA}
}
\maketitle
\begin{abstract}
We rigorously rule out the appearance of multiple domain walls between
ground states in $2D$ Edwards-Anderson Ising spin glasses (with periodic
boundary conditions and, e.g., Gaussian couplings).  This supports the
conjecture that there is only a single pair of ground states in these
models.
\end{abstract}
\small
\normalsize

A fundamental problem in spin glass physics is the multiplicity of
infinite-volume ground states in finite-dimensional short-ranged systems,
such as the Edwards-Anderson (EA) \cite{EA} Ising spin glass.  In $1D$,
there is no frustration and only a single pair of (spin-reversed) ground
states.  In the mean-field Sherrington-Kirkpatrick (SK) model \cite{SK},
there are presumed to be (in some suitably defined sense) infinitely many
ground state pairs (GSP's) \cite{Parisi79}.  One conjecture, in analogy
with the SK model, is that finite $D$ realistic models with frustration
have infinitely many GSP's; for a review, see \cite{MPV,BY}.  A different
conjecture, based on droplet-scaling theories \cite{Mac,BM,FH86}, is that
there is only a single GSP in all finite $D$.  In $2D$ and 
$3D$, the latter scenario
has received support from recent simulations, some
\cite{Middleton,PY} based on ``chaotic size
dependence'' \cite{NS92} and some \cite{Hartmann} using other techniques.

In this paper, we provide a significant analytic step towards a
resolution of this problem in $2D$, by ruling out the presence of multiple
domain walls between ground states.  We anticipate that the ideas and
techniques introduced here will ultimately yield a solution to the problem
of ground state multiplicity in two dimensions, and that at least some of
them may prove to be useful in higher dimensions as well.  Though our
result is more general, we confine our attention to the nearest-neighbor EA
Ising spin glass, with Hamiltonian
\begin{equation}
\label{eq:EA}
{\cal H}_{\cal J}(\sigma)= -\sum_{\langle x,y\rangle} J_{xy} \sigma_x \sigma_y\quad ,
\end{equation}
where ${\cal J}$ denotes a specific realization of the couplings $J_{xy}$,
the spins $\sigma_x=\pm 1$ and the sum is over nearest-neighbor pairs
$\langle x,y\rangle$ only, with the sites $x,y$ on the square lattice ${\bf
Z}^2$.  The $J_{xy}$'s are independently chosen from a mean zero Gaussian
(or any other symmetric, continuous distribution with unbounded support)
and the overall disorder measure is denoted $\nu ({\cal J})$.

A ground state is an infinite-volume spin configuration whose energy
(governed by Eq.~(\ref{eq:EA})) cannot be lowered by flipping any finite
subset of spins.  That is, all ground state spin configurations must
satisfy the constraint
\begin{equation}
\label{eq:loop}
\sum_{\langle x,y\rangle\in {\cal C}}J_{xy}\sigma_x \sigma_y \ge 0
\end{equation}
along any closed loop ${\cal C}$ in the dual lattice.  In any $L\times L$
square $S_L$ (centered at the origin) with, e.g., periodic b.c.'s, there is
(with probability one) only a single finite-volume GSP (the spin
configurations of lowest energy subject to the b.c.).  An infinite-volume
ground state can be understood as a limit of finite-volume ones: consider
the ground state ${\sigma}^{(L_0,L)}$ inside any given $S_{L_0}$, but with
b.c.'s imposed on $S_{L}$ and $L\gg L_0$.  An infinite-volume ground state
(satisfying Eq.~(\ref{eq:loop})) is generated whenever, for each (fixed)
$L_0$, ${\sigma}^{(L_0,L)}$ converges to a limit as $L\to\infty$ (for some
sequence of b.c.'s, which {\it may\/} depend on the coupling realization).
If many infinite-volume GSP's exist, then a sequence as $L\to\infty$ of
finite-volume GSP's with coupling-{\it independent\/} b.c.'s will generally
{\it not\/} converge to a single limit (i.e., ${\sigma}^{(L_0,L)}$
continually changes as $L\to\infty$), a phenomenon we call {\it chaotic
size dependence\/}
\cite{NS92}.  So a numerical signal of the existence of many ground states
is that the GSP in $S_L$ with periodic b.c.'s varies chaotically as $L$
changes \cite{Middleton,PY,NS92}.

It is important to distinguish between two types of multiplicity.  The {\it
symmetric difference\/} $\alpha \Delta \beta$ between two GSP's $\alpha$
and $\beta$ is the set of all couplings that are satisfied in one and not
the other.  A {\it domain wall\/} (always defined relative to two GSP's) is
a cluster (in the dual lattice) of the couplings satisfied in one but not
the other state.  So $\alpha \Delta \beta$ is the union of all of their
domain walls, and may consist of a single one or many.  Two distinct GSP's
are {\it incongruent\/} \cite{HF87} if $\alpha \Delta \beta$ has
nonvanishing density in the set of all bonds; otherwise the two are {\it
regionally congruent\/}.  Incongruent GSP's can in principle have one or
more positive density domain walls, or instead infinitely many of zero
density.

If there are multiple GSP's, the interesting, and physically relevant,
situation is the existence of incongruent states.  Regional congruence is
of mathematical interest, but to see it would require a choice of b.c.'s
carefully conditioned on the coupling realization ${\cal J}$.  It is not
currently known how to choose such b.c.'s.  Numerical treatments that look
for multiple GSP's implicitly search for incongruent ground states, and it
is the question of their existence and nature in $2D$ that we treat here.

To state our result precisely, we introduce the concept of a {\it
metastate\/}.  For spin glasses, this was proposed in the context of low
temperature states for large finite volumes \cite{NS96b} (and shown to be
equivalent to an earlier construct of Aizenman and Wehr \cite{AW}), and its
properties were further analyzed in \cite{NS97,NS98}.  In the current
context, a (periodic b.c.) metastate is a measure on GSP's constructed via
an infinite sequence of squares $S_L$, with both the $L$'s and the
(periodic) b.c.'s coupling-independent.  Roughly speaking, the metastate
here provides the probability (as $L \to \infty$) of various GSP's
appearing inside any fixed $S_{L_0}$.  It is believed (but not proved) that
different sequences of $L$'s yield the same (periodic b.c.) metastate.

If there are infinitely many (incongruent) GSP's, a metastate should be
dispersed over them, giving their relative likelihood of appearance in
typical large volumes.  If there is no incongruence, the metastate would be
unique and supported on a single GSP, and that GSP will appear in most (i.e., a
fraction one) of the $S_L$'s \cite{fraction}.

We now state the main result of this paper.  It shows that if more than a
single GSP is present in the periodic b.c.~metastates, then two distinct GSP's
cannot differ by more than a {\it single\/} domain wall.  After we present
the proof of this statement, we will discuss why this result supports the
existence of only a single GSP in $2D$.

{\it Theorem.\/} In the $2D$ EA Ising spin glass with
Hamiltonian~(\ref{eq:EA}) and couplings as specified earlier, two
infinite-volume GSP's chosen from the periodic b.c.~metastates 
are either the same or else differ by a
single, non-self-intersecting domain wall, which has positive density.

We sketch the proof of this theorem in several steps; a full presentation
will be given elsewhere \cite{NSinprep}.  First, some elementary properties
of (zero-temperature) domain walls:

{\it Lemma 1.\/} A $2D$ domain wall is infinite and contains no loops or
dangling ends.

{\it Proof\/}. A domain wall between two spin configurations is a boundary
separating regions of agreement from disagreement and thus cannot have
dangling ends.  To rule out loops, note that the sum
$\sum_{<xy>}J_{xy}\sigma_x\sigma_y$ along any such loop must have opposite
signs in the two GSP's, violating Eq.~(\ref{eq:loop}), unless the sum
vanishes. But this occurs with probability zero because the couplings are
chosen independently from a continuous distribution.

We now construct a periodic b.c.~metastate $\kappa_{\cal J}$, which will
provide a measure on the domain walls between GSP's (that appear in
$\kappa_{\cal J}$).  As in construction II of \cite{NS96c} (but at zero
temperature), consider for each square $S_L$, two sets of variables, the
couplings ${\cal J}^{(L)}$ (chosen from, e.g., the Gaussian distribution)
and the bond variables $\sigma_x^{(L)}\sigma_y^{(L)}$ for the GSP $\pm
\sigma^{(L)}$.  Consider fixed sets of both random variables as
$L\to\infty$; by compactness, there exists a subset of $L$'s along which
the joint distribution converges to a translation-invariant infinite-volume
(joint) measure.  This limit distribution is supported on ${\cal J}$'s that
arise from $\nu$, the usual independent (e.g., Gaussian) distribution on
the couplings, and the conditional (on ${\cal J}$) distribution
$\kappa_{\cal J}$ is supported on (infinite-volume) GSP's for that ${\cal
J}$.

A metastate $\kappa_{\cal J}$ yields a measure ${\cal D}_{\cal J}$ on
domain walls.  This is done by taking two (replica) GSP's from
$\kappa_{\cal J}$ to obtain a configuration of (unions of) domain walls
(i.e., the set of domain walls one would see from two GSP's chosen randomly
from $\kappa_{\cal J}$).  If one then integrates out the couplings, one is
left with a translation-invariant measure ${\cal D}$ on the domain wall
configurations themselves.

This leads to
important percolation-theoretic features of domain walls between GSP's in
$\kappa_{\cal J}$.  
Some of these \cite{BK1} are stated in the following: 

{\it Lemma 2.\/} Distinct $2D$ GSP's $\alpha$ and $\beta$ from
$\kappa_{\cal J}$ must (with probability one) be incongruent and the domain
walls of their symmetric difference $\alpha \Delta \beta$ must be
non-intersecting, non-branching paths, that together divide ${\bf Z}^2$
into infinite strips and/or half-spaces.

{\it Proof.\/} This lemma, from \cite{BK1}, uses a technique introduced in
\cite{BK2}.  First we note that by the translation-invariance of ${\cal
D}$, any ``geometrically defined event'', e.g., that a bond belongs to a
domain wall, either occurs nowhere or else occurs with strictly positive
density. This immediately yields incongruence. Suppose now that an
intersection/branching occurs at some site $z$ (in the dual lattice).  Then
there are at least three (actually four) {\it infinite\/} paths in
$\alpha \Delta \beta$ that start from $z$, and they cannot intersect in
another place, because that would form a loop, violating Lemma 1.  But then
translation-invariance implies a positive density of such $z$'s.  The
tree-like structure of $\alpha \Delta \beta$ implies that in a square with
$p$ such $z$'s, the number of distinct such paths crossing its boundary is
at least proportional to $p$.  Since $p$ scales like $L^2$, there is a
contradiction as $L\to\infty$, because the number of distinct paths cannot
be larger than the perimeter, which scales like $L$.  Similar arguments
complete the proof.

The picture we now have for $\alpha \Delta \beta$ is a union of one or more
infinite domain walls (each of which divides the plane into two infinite
disjoint parts) that neither branch, intersect, nor form loops, and that
mostly remain within $O(1)$ distance from one another.  We now begin a
lengthy argument to show that there in fact {\it cannot\/} be more than a
single domain wall.  The first step is to introduce the notion of a
``rung'' between adjacent domain walls.

A rung ${\cal R}$ in $\alpha \Delta \beta$ is a path of bonds in the dual
lattice connecting two distinct domain walls, and with only the first and
last sites in ${\cal R}$ on any domain wall.  So each of the couplings in
${\cal R}$ is satisfied in both $\alpha$ and $\beta$ or unsatisfied in
both.  The energy $E_{\cal R}$ of ${\cal R}$ is defined to be

\begin{equation}
\label{eq:rung}
E_{\cal R} = \sum_{<xy>\in{\cal R}} J_{xy}\sigma_x\sigma_y \quad ,
\end{equation}
with $\sigma_x\sigma_y$ taken from $\alpha$ (or equivalently, $\beta$).  It
must be that $E_{\cal R}>0$ (with probability one) for the following
reason.  Suppose that a rung could be found with negative energy; by
translation-invariance (and arguments somewhat like those used for Lemma 2),
there would then be an infinite set of rungs with negative energy
connecting some two domain walls.  Consider the ``rectangle'' that is
bounded by two such rungs and the connecting domain wall pieces.  The sum
of $J_{xy}\sigma_x\sigma_y$ along the couplings in the two domain wall
pieces would be positive in one of $\alpha,\,\beta$ and negative in the
other; hence, the loop formed by the boundary of this rectangle would
violate Eq.~(\ref{eq:loop}) in $\alpha$ or $\beta$, leading to a
contradiction.

However, we can impose a more serious constraint on $E_{\cal R}$; namely
that it must be bounded {\it away\/} from zero for all ${\cal R}$ between
two fixed domain walls.  To explain this, we first consider a single
arbitrary bond $b$, an $S_L$ large enough to contain $b$, a coupling
realization ${\cal J}^{(L)}$ and the corresponding GSP $\alpha^{(L)}$.  Now
let $J_b$ vary with all other couplings fixed.  It is easy to see that
there will be a transition value $K_b^{(L)}$ (which is a function of all
the couplings in ${\cal J}^{(L)}$ {\it except\/} $J_b$) beyond which
$\alpha^{(L)}$ ceases to have minimum energy and is replaced by some
$\alpha^{b,(L)}$, related to $\alpha^{(L)}$ by a droplet flip. The
symmetric difference $\alpha^{(L)}\Delta \alpha^{b,(L)}$ consists of a
domain wall (the boundary of the droplet) passing through $b$ with exactly
zero total energy when $J_b = K_b^{(L)}$.  The droplet boundary may or may
not reach the boundary of $S_L$.  In other words, as $J_b$ varies from
$-\infty$ to $+\infty$, there are exactly two GSP's ($\alpha^{(L)}$ and
$\alpha^{b,(L)}$) that appear, one when $J_b$ is below $K_b^{(L)}$ and one
when it is above.

What happens when $L\to\infty$?  As in the construction of metastates, we
obtain a translation-invariant infinite-volume joint probability
distribution on ${\cal J}$ (the couplings $J_b$), $\alpha$ (a GSP for
${\cal J}$), ${\cal K}$ (transition values $K_b$ for ${\cal J},\,\alpha$)
and $\alpha^*$ ($\alpha^b$'s for ${\cal J},\,\alpha,\,{\cal K}$).  In this
limit: ${\cal J}$ is chosen from the usual disorder distribution $\nu$,
then $\alpha$ from the metastate $\kappa_{\cal J}$ and finally ${\cal K}$
and $\alpha^*$ from some measure $\kappa_{{\cal J},\alpha}$.  The symmetric
difference $\alpha \Delta \alpha^b$ may consist of a single finite loop or
else of one or more infinite disconnected paths, but in all cases some part
must pass through $b$. The lack of dependence of $K_b^{(L)}$ on $J_b$
implies that even after $L\to\infty$, {\it $K_b$ and $J_b$ are independent
random variables\/}; this independence leads to the next two lemmas.

{\it Lemma 3.\/}  With probability one, no coupling $J_b$ is
exactly at its 
transition value $K_b$.

{\it Proof.\/} From the independence of $J_b$ and $K_b$, and the continuity
of the distribution of $J_b$, it follows that there is probability zero
that $J_b - K_b =0$, much like in the proof of Lemma 1.

{\it Lemma 4.\/} The rung energies $E_{{\cal R}'}$
between two fixed (adjacent) domain walls
cannot be arbitrarily small; i.e., there is zero probability
that $E'$, the infimum of all such $E_{{\cal R}'}$'s, will be zero. 

{\it Proof.\/} Were this not so, there would be (by translation-invariance
arguments) an infinite set of rungs ${\cal R}'$ with $E_{{\cal R}'} <
\epsilon$, for any $\epsilon>0$.  That implies (by the ``rectangular''
construction below Eq.~(\ref{eq:rung})) that each $J_b$ along the two
domain walls would be at the transition value $K_b$, either for $\alpha$ or
for $\beta$, violating Lemma 3.

The next lemma relates the location of the droplet boundary, $\alpha \Delta
\alpha^a$, when $\alpha^a$ replaces $\alpha$, to the ``flexibility'' of
$a$.  The flexibility $F_a$ of a bond $a$ (in a $({\cal J},\alpha,{\cal
K},\alpha^*)$ configuration) is defined as $|J_a - K_a|$; the larger the
flexibility, the more stable is $\alpha$ under changes of $J_a$.

{\it Lemma 5\/.}  If $F_b > F_a$, then there is zero probability
that $\alpha \Delta \alpha^a$ passes through $b$.

{\it Proof.\/} For finite $L$, this is an elementary consequence of the
fact that for $e=a$ or $b$, $F_e^{(L)}\equiv |J_e - K_e^{(L)}|$ is the
minimum, over all droplets whose boundary passes through $e$, of the
droplet flip energy cost.  After $L \to \infty$, such a characterization of
$F_e$ may not survive, but what does survive is that $\alpha \Delta
\alpha^a$ does not go through $b$.

The next lemma completes our proof that for GSP's $\alpha$ and $\beta$
chosen from $\kappa_{\cal J}$, $\alpha \Delta \beta$ cannot consist of more
than a single domain wall, since otherwise there would be an immediate
contradiction with Lemma 4. For the proof, we need the notion of
``super-satisfied''.  It is easy to see that a coupling $J_{xy}$ is
satisfied in every ground state if $|J_{xy}|>$min$\{M_x,M_y\}$, where $M_x$
is the sum of the three other coupling magnitudes $|J_{xz}|$ touching $x$,
and $M_y$ is defined similarly.  Such a coupling $J_{xy}$, called
super-satisfied, clearly cannot be part of any domain wall.

{\it Lemma 6\/.}  There is zero probability that $E'>0$.

{\it Proof.\/} Suppose $E'>0$ (with positive probability); we show this
leads to a contradiction.  First we find, as in Fig.~1, a rung ${\cal R}$
with $E_{\cal R}-E'=\delta$ strictly less than the flexibility values (for both
$\alpha$ and $\beta$) of two couplings $b_1,b_2$ along the ``left'' of the
two domain walls, $b_1$ ``above'' and $b_2$ ``below'' the rung. Such an
${\cal R}$, $b_1$ and $b_2$ must exist by Lemma 3 (and
translation-invariance arguments).

\begin{figure}
\centerline{\epsfig{file=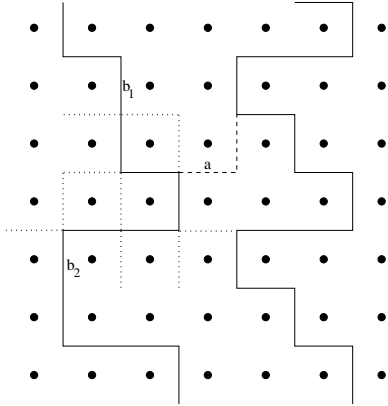,width=2.0in}}
\vspace{0.2in}
\caption{A rung ${\cal R}$ with $E_{\cal R} = E' +\delta$.  The dots are
sites in ${\bf Z}^2$, and couplings are drawn in the dual lattice.  Two
domain walls are solid lines and ${\cal R}$ is the dashed line.  The
couplings $b_1$ and $b_2$ have flexibility $>\delta$.  The ten dotted line
couplings are super-satisfied.}
\label{fig:bounce}
\end{figure}

But we also want a situation, as in Fig.~1, where all the (dual lattice)
non-domain-wall couplings that touch the left domain wall between $b_1$ and
$b_2$ (other than the first coupling $J_a$ in ${\cal R}$) are
super-satisfied, and remain so regardless of changes of $J_a$.  How do we
know that such a situation will occur (with non-zero probability)?  If
necessary, one can first adjust the signs and then increase the magnitudes
(in an appropriate order) of these (ten) couplings, so that they first
become satisfied and then super-satisfied. This can be done in an
``allowed'' way because of our assumption that the distribution of
individual couplings has unbounded support. Also, this can be done without
causing a replacement of either $\alpha$ or $\beta$, without changing
$E_{\cal R}$, without decreasing any other $E_{{\cal R}'}$ and without
decreasing the flexibilities of $b_1$ or $b_2$. Starting from a positive
probability event, such an (allowed) change of finitely many couplings in
${\cal J}$ yields an event which still has non-zero probability.

Next, suppose we move $J_a$ toward its transition value $K_a$ by an amount
slightly greater than $\delta$.  The geometry (of Fig.~1) and Lemma 5
forbid the replacement of either $\alpha$ or $\beta$, because it is
impossible, under the conditions given, for $\alpha \Delta \alpha^a$ or
$\beta \Delta \beta^a$ to connect to the left end of bond $a$.  But this
move reduces $E_{\cal R}$ below $E_{{\cal R}'}$ for any ${\cal R}'$ not
containing $a$, contradicting translation-invariance.

This completes the proof of the theorem: if distinct $\alpha, \beta$ occur,
they differ by at most a {\it single\/} domain wall.  Although this does
not yet rule out many ground states in the $2D$ periodic b.c.~metastate, it
greatly simplifies the problem by ruling out all but one possibility about
how GSP's may differ.

We expect, though, that these single domain walls do {\it not\/} exist.
There are reasonable arguments and conjectures indicating that this is so,
and that even if they do exist, it remains unlikely that there exists an
infinite multiplicity of states.  We will discuss these in turn.

First, we note that although, for technical reasons, we have not 
extended our proof to rule out single domain walls, our previous results
indicate that it is natural to expect that the ``pseudo-rungs'' that
connect sections of the domain wall that are close in Euclidean distance,
but greatly separated in distance along the domain wall, can have
arbitrarily low (positive) energies.  If these ``pseudo-rungs'' also
connect arbitrarily large pieces of the domain wall containing some fixed
bond (and we emphasize that these properties are not yet rigorously
proved), then single domain walls would be ruled out in a similar manner as
above.  The consequence would be that the periodic b.c.~metastate in the
$2D$ EA Ising spin glass with Gaussian couplings is supported on a single
GSP.

In the unlikely event that single positive-density domain walls {\it do\/}
appear, our theorem could still rule out an {\it infinite\/}
multiplicity of GSP's in $2D$.  
This would be a consequence of the following conjecture (which presents an
interesting problem in the topology of random curves):

{\it Conjecture:\/} There exists no translation-invariant measure on
infinite sequences $(a_1,a_2,\dots)$ of distinct bond configurations on
${\bf Z}^2$ such that each $a_i$ and each $a_i \Delta a_j$ is a single,
doubly-infinite, self-avoiding path.

The above conjecture, if true, would rule out the presence of infinitely
many distinct GSP's $\alpha_0,\alpha_1,\dots$ (in one or more metastates
for a given ${\cal J}$) since taking $a_i = \alpha_0 \Delta \alpha_i$ would
contradict the conjecture.

These considerations, taken together, make it appear unlikely that an
infinite multiplicity of GSP's, constructed from periodic (or antiperiodic
\cite{NS98}) boundary conditions, can exist for the $2D$ EA Ising spin
glass with Gaussian (or similar) couplings.

\medskip

\renewcommand{\baselinestretch}{1.0}


\begin{thebibliography}{10}
\bibitem{fraction}
Domain walls that fluctuate away from the origin would not 
appear in the metastate.
\bibitem{NSinprep} C.M.~Newman and D.L.~Stein, in preparation.  Our sketch
is for GSP's from a single $\kappa_{\cal J}$, but the same arguments apply
if they are from different metastates.
\end{thebibliography}
\end{document}